\documentclass[a4paper,fleqn,usenatbib]{mnras}
\usepackage[T1]{fontenc}
\usepackage{ae,aecompl}
\usepackage{graphicx}
\usepackage{amssymb}
\usepackage{amsmath}
\usepackage{mathptmx}
\usepackage{epstopdf}
\usepackage{booktabs}
\usepackage{float}
\usepackage{subfig}
\usepackage{epsfig,color,ulem}
\usepackage{tabularx}

\newcommand{\A}{{\it A}}
\newcommand{\B}{{\it B}}
\newcommand{\C}{{\it C}}

\newcommand{\cm}{\,{\rm cm}}

\newcommand{\erg}{\,{\rm erg}}

\def\gsim{ \lower .75ex \hbox{$\sim$} \llap{\raise .27ex \hbox{$>$}} }
\def\lsim{ \lower .75ex\hbox{$\sim$} \llap{\raise .27ex \hbox{$<$}} }
\def\app#1#2{%
	\mathrel{%
		\setbox0=\hbox{$#1\sim$}%
		\setbox2=\hbox{%
			\rlap{\hbox{$#1\propto$}}%
			\lower1.1\ht0\box0%
		}%
		\raise0.25\ht2\box2%
	}%
}

\title[Detectability of neutron star merger afterglows]{Detectability of neutron star merger afterglows}

	\author[Gottlieb, Nakar \& Piran]{
		Ore Gottlieb$^{1}$\thanks{oregottlieb@mail.tau.ac.il},
		Ehud Nakar$^{1}$,
		Tsvi Piran$^{2}$
		\\
		$^{1}${The Raymond and Beverly Sackler School of Physics and
			Astronomy, Tel Aviv University, Tel Aviv 69978, Israel}\\
		$^{2}${Racah Institute of Physics, The Hebrew University of
			Jerusalem, Jerusalem 91904, Israel}
}
\pubyear{2019}
\begin{document}
	\label{firstpage}
	\pagerange{\pageref{firstpage}--\pageref{lastpage}}
	\maketitle	
	\begin{abstract}

	VLBI and JVLA observations revealed that GW170817 involved a narrow jet ($ \theta_j \approx 4^\circ $) that dominated the afterglow peak at our viewing angle, $ \theta_{\rm obs} \approx 20^\circ $.
	This implies that {\it at the time of the afterglow peak}, the observed signal behaved like an afterglow of a top-hat jet seen at $ \theta_{\rm obs} \gg \theta_j $, and it can be modeled by analytic expressions that describe such jets.
	We use a set of numerical simulations to calibrate these analytic relations and obtain generic equations for the peak time and flux of such an afterglow as seen from various observing angles.
	Using the calibrated equations and the estimated parameters of GW170817, we estimate the detectability of afterglows from future double neutron star mergers during the Advanced LIGO/Virgo observation run O3.
	GW170817 took place at a relatively low-density environment. Afterglows of similar events will be detectable only at small viewing angles, $ \theta_{\rm obs} \lesssim 20^\circ $, and only $\sim 20\% $ of the GW detections of these events will be accompanied by a detectable afterglow. At higher densities, more typical to sGRB sites, up to $ 70\% $ of the GW detections are expected to be followed by a detectable afterglow, typically at $ \theta_{\rm obs} \sim 30^\circ $. We also provide the latest time one should expect an afterglow detection. We find that for typical parameters, if the jet emission had not been detected within about a year after the merger, it is unlikely to be ever detected. 
	
	\end{abstract}
	\begin{keywords}
		{gamma-ray burst: short | stars: neutron | gravitational waves | methods: analytical}
	\end{keywords}

\section{Introduction}
\label{sec:introduction}	

GW170817, the first gravitational waves (GW) signal from a binary neutron star merger  was detected by advanced LIGO/Virgo on August 17 2017. It was accompanied by several electromagnetic (EM) counterparts, including 
an unprecedented faint $ \gamma $-rays signal \citep{Goldstein2017,Savchenko2017}, a UV/optical/IR macronova/kilonova signal (hereafter called macronova),
\citep{Andreoni2017,Arcavi2017,Arcavi2017a,Buckley2017,Chornock2017,Coulter2017,Cowperthwaite2017,Diaz2017,Drout2017,Evans2017,Hu2017,Kasliwal2017,Kilpatrick2017,Lipunov2017,McCully2017,Nicholl2017,Pian2017,Pozanenko2018,Shappee2017,Smartt2017,Soares-Santos2017,Tanvir2017,Tominaga2017,Utsumi2017,Valenti2017,Villar2017,Villar2018} and a non thermal X-ray, optical and radio afterglow \citep{Hallinan2017,Dobie2018,Mooley2018a,Margutti2017,Margutti2018,Lyman2018,Troja2017,Troja2018,Haggard2017,Alexander2017,Alexander2018,DAvanzo2018,Mooley2018c,Nynka2018,Ruan2018,Lamb2019}.

It was clear right from the beginning \citep{Kasliwal2017} that the weak prompt $\gamma$-ray signal was not  a regular GRB seen off-axis (namely that the same  $\gamma$-rays that we observed were not seen by an alien on-axis observer as a regular GRB; see also \citealt{Matsumoto2018}).  
A plausible origin of this $\gamma$-ray signal is a shock breakout from the cocoon which is formed when a jet propagates through the ejecta \citep{Kasliwal2017,Bromberg2017,Gottlieb2018b,Lazzati2018,Mooley2018a,Nakar2018,Pozanenko2018,Beloborodov2018}. As the cocoon shock breakout can take place regardless of whether the jet breaks out or not, the observed gamma-rays did not answer the question of whether the event involved a jet that emerged and produced a regular sGRB pointing elsewhere. The afterglow which rose gradually over the first few months did not provide an immediate answer as well. However, \cite{Mooley2018b} show that the fast decline that followed the peak combined with VLBI observations \citep[these VLBI observations are consistent with those of][that support these conclusion]{Ghirlanda2019} revealed that GW170817 was accompanied by a narrow ($ \theta_j \approx 4^\circ $) and powerful ($ E_{iso} \approx 3\times 10^{52}\erg $) jet that pointed  $ \theta_{\rm obs} \approx {20^\circ}^{+9^\circ}_{-5^\circ} $ relative to us and dominated the peak of the afterglow emission and the subsequent decline
(see also \citealt{Lazzati2018,Alexander2018,Mooley2018c,vanEerten2018,Lamb2019}).

Toward the upcoming Advanced LIGO/Virgo observation run O3, it is interesting to explore the detectability of future joint GW/EM detections from binary neutron star mergers.
At very small viewing angles the regular GRB will be easily detectable with a detection horizon much larger than the expected O3 GW detection horizon $\sim 200 $Mpc. At larger viewing angles the cocoon shock breakout, that was most likely the source of the observed $\gamma$-rays in GRB170817a, may still be observable. However this signal is expected to decay strongly with viewing angle. The same is true for other sources of gamma-ray emission discussed in the literature (e.g., off-axis emission from the jet core). Thus, given that the detection horizon of  GRB170817a ($\sim 60$Mpc), we expect that it will be nearly impossible to detect a gamma-ray signal from $\theta_{\rm obs} \gtrsim 20^\circ $ at the O3 GW horizon.
Thus, given the small solid angle at which the jet can be detected directly (see e.g. \citealt{Beniamini2019a,Beniamini2019b}), it is of outmost interest to capture the afterglow in other events as well, as this may be the only direct observation of the relativistic outflow component.

Here we utilize the lessons learned from GW170817 and explore the detectability of the radio (and X-ray) afterglow in future binary neutron star merger events whose GW signal will be detected by Advanced LIGO/Virgo in O3.
The search for the afterglow depends naturally on whether there is a precise localization of the merger via the macronova signal, which is expected to be isotropic. When such a localization is available (as in GW170817, which had a bright macronova signal with an AB magnitude of about $17$), sensitive radio telescopes and X-ray satellites are pointed towards this location in search for the afterglow.
If the early UV/optical component seen in GW170817 is generic, its magnitude for a source at 200 Mpc will be about $21$, allowing a localization as long as the event does not take place behind the Sun. The situation is quite different if the UV/optical component is missing and we will have to search for the IR component, which with current facilities will be much more difficult to detect.

In this paper the canonical radio afterglow sensitivity limit is taken under the assumption that most of the merger events will have an accurate macronova localization. 
Sensitivity for blind afterglow searches over the GW detectors' detection area will be lower although it may still be sensitive enough to detect the afterglow \citep{Dobie2019}. 
Considering, instead, the detectability of a blind survey is straightforward using the formulae we provide. 
We focus on observers at angles much larger than the jet's core, $\theta_{\rm{obs}} \gg \theta_j $, where the very bright multi-wavelength emission from the jet is not detectable.
Clearly, the detectability depends on the strongest signal and hence on the peak of the afterglow. The important parameters for a detection are, therefore, the peak flux and the peak time of the afterglow light curves. We focus on radio afterglow detectability, as it is most robustly predicted  and easiest to detect. Thus,  when referring to afterglow in the rest of the text we mean radio-afterglow. However, the result can be easily scaled to X-ray or optical observations, as discussed in 
\S \ref{sec:optical_xray}.

\citet{Nakar2002} and \citet{Totani2002} have derived analytic relations for the peak flux and time for off-axis afterglows of top-hat jets. Remarkably, it turns out that these results are valid  for a wide range of jets, regardless of their angular structure.  In the first part of this paper we discuss the applicability of these analytic relations to a large variety of GRBs, including ones from double NS mergers. We then calibrate and verify it with a set of numerical simulations (see also \citealt{Granot2018}). We show in $\S $ \ref{sec:analytic_fit} that if the jet energy is large enough and dominates the peak afterglow, then both the peak flux and the peak time do not depend on the detailed angular structure, and the model for top-hat jets is applicable. 
In the second part of this paper ($\S $ \ref{sec:detectability}) we use the calibrated analytic equations to explore the detection rates and detection horizon of future afterglows from double NS mergers during the upcoming run O3. We begin with considering events with the same parameters as those inferred to GW170817, then we generalize our results to events with other circum-merger densities and jet's energies, as these parameters are expected to be different from one merger to another. In $ \S $ \ref{sec:conclusions} we summarize and conclude.

\section{The peak time and flux of jet dominated afterglows}
\label{sec:analytic_fit}
We define  the jet core  as the region  inside $\theta_j$ in which most of the jet's energy  is contained.
 A jet {that is} observed at $\theta_{obs} \gg \theta_j$  dominates the  afterglow peak  if the jet core is more energetic than the cocoon that surrounds it. This situation is {expected} when the engine launching the jet continues to work for a significant amount of time after the jet breaks out from the surrounding matter. In such systems the observed emission {\it at the peak} behaves like an off-axis (orphan) afterglow {of} a top-hat jet, regardless of the exact jet structure. The latter was explored analytically in the past
\citep{Granot2002,Nakar2002,Totani2002}.

We show here that  the analytic top-hat relations are generally valid  for different GRBs viewed off-axis  that satisfy the above mentioned conditions\footnote{In principle there are systems in which these criteria may be violated. For example, the contribution of the flux from the cocoon can be  comparable to that of the jet at the time of the peak.}:
(a) The jet's afterglow dominate the peak of the light curve.
 (b) The jet core opening angle, $ \theta_j $, that is much smaller than the  viewing angles ($ \theta_j \ll \theta_{\rm{obs}} $).
In particular, in GW170817 the light curve decline following the peak indicated that both
conditions have almost certainly been fulfilled.  

For  $ \theta_{\rm{obs}} \gg \theta_j $, Eqs. 5 and 10 in \citet{Nakar2002} provide the afterglow peak flux and peak time as functions of the system's parameters for frequencies above the self absorption frequency and below the cooling frequency.
In particular, for a given set of parameters these equations predict that the peak flux and peak time satisfy $ F_{\nu,p} \propto \theta_{\rm obs}^{-2p} $ and $ t_p \propto (\theta_{\rm obs}-\theta_j)^2 $, where $ p $ is the electrons' distribution power-law index.

We compare the analytic relations and calibrate them using  four hydrodynamic simulations of jet propagation, the corresponding blast wave and the resulting afterglow emission. The configurations studied include a simple  top-hat jet as considered in the analytic models, and three configurations that resulted from a jet propagating within the ejecta \citep[simulations $ \A,\B,\C $ in][]{Mooley2018b}. These configurations have rather complicated and different angular structures. All configurations satisfy the two criteria listed above. 
Following the relativistic hydrodynamic simulations we post-process the results to obtain the afterglow light curves as seen by various observers (see Appendix \ref{sec:post_process} for the numerical methods). 
\begin{figure}
	\centering
	\includegraphics[width=0.5\textwidth]{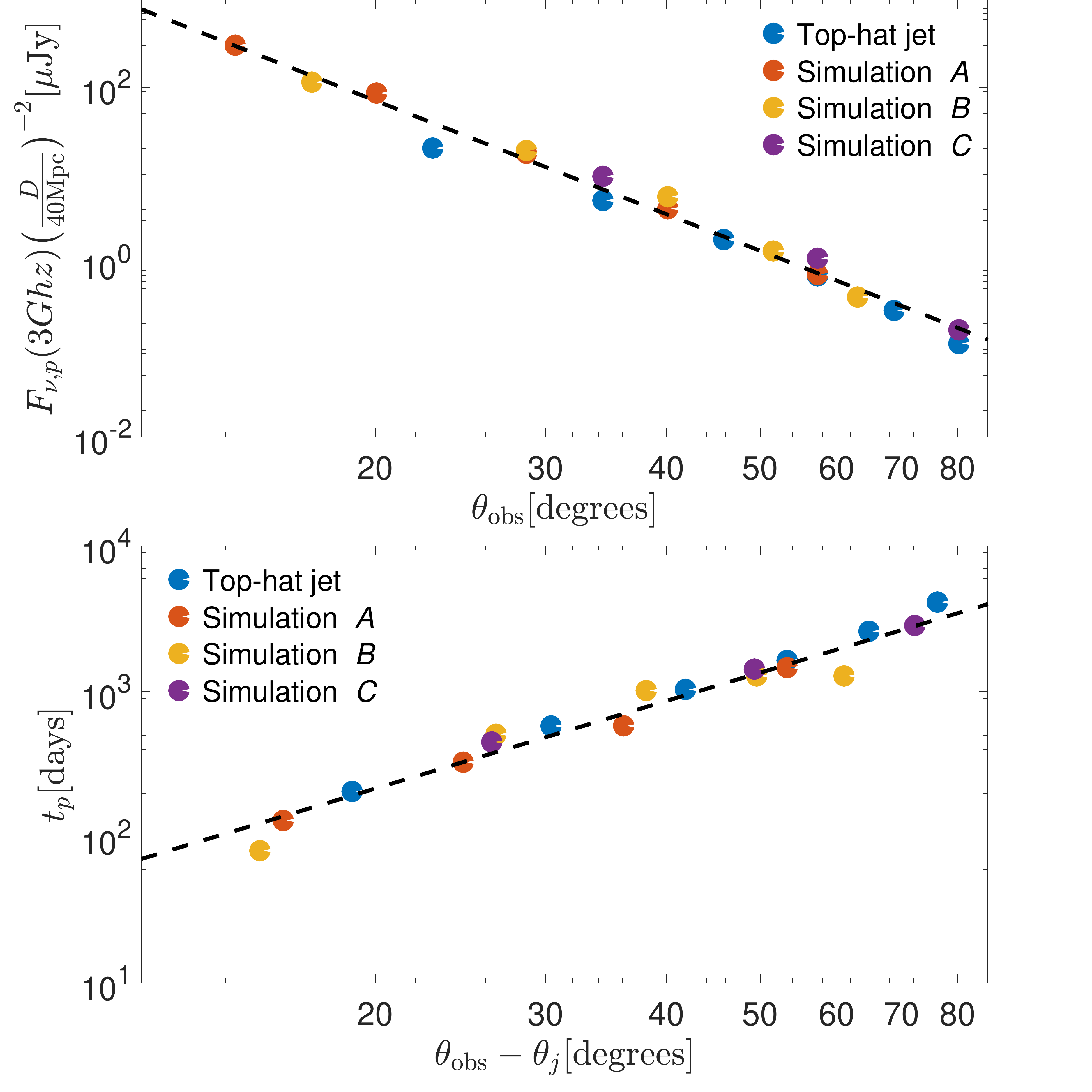}
	\caption[Light curves for different observers]{
		The peak fluxes (top) and peak times (bottom) at various viewing angles in the different numerical simulations. All simulations share the same set of parameters, $ E = 10^{50}\erg,~ n = 10^{-3}\cm^{-3},~ \epsilon_B= 6\times 10^{-5} $ and $ \epsilon_e = 0.1 $.
		The black dashed lines depict Eqs. \ref{eq:peak_magn} and \ref{eq:peak_timen}, using the canonical parameters and $ p = 2.16 $.
		Both analytic relations agree well with the numerical data.
	}
	\label{fig:analytic}
\end{figure}

Figure \ref{fig:analytic} depicts the peak flux and peak time at different angles as obtained in the simulations.  
The black dashed curve represents the analytic relations $ F_{\nu,p} \propto \theta_{\rm obs}^{-2p} $ (for $ p = 2.16 $)  and $ t_p \propto (\theta_{\rm obs}-\theta_j)^2 $. The normalization of the curve is chosen to fit the numerical results.
We find, as expected, that not only do all numerical models follow the functional dependence at different viewing angles, but they also share the same normalization. 
When using  an electron energy power-law index $ p = 2.5 $ for the top-hat simulation, we find the functional dependence  on $p$ is in a very good agreement with the one given in \citet{Nakar2002}, denoted as $ g(p) $ below.

Using  the numerically determined normalization we calibrate the analytic equations (5 $ \& $ 10 in \citealt{Nakar2002}) which now take the form:
\begin{eqnarray}
%\begin{equation}
\label{eq:peak_magn}
F_{\nu,p}&=& 90 \bigg\{ \frac{g(p)}{g(2.16)} 
\bigg(\frac{\epsilon_e}{0.1}\bigg)^{p-1}
\bigg( \frac{\epsilon_B}{6\times 10^{-5}}\bigg)^{\frac{p+1}{4}} 
\nonumber \bigg\} 
 \\ 
&\times& \bigg[ \bigg(  \frac{E}{10^{50}{\rm erg}} \bigg) 
\bigg( \frac{n}{10^{-3}\cm^{-3}}  \bigg)^{\frac{p+1}{4}} 
\bigg] 
 \\
&\times&   \bigg(\frac{\nu}{3\rm{GHz}}\bigg)^{\frac{1-p}{2}}\bigg(\frac{\theta_{\rm{obs}}}{20^\circ}\bigg)^{-2p}\bigg(\frac{D}{40\rm{Mpc}}\bigg)^{-2}  \mu\rm{Jy} \ ,
\nonumber
\end{eqnarray}
and
\begin{equation}\label{eq:peak_timen}
t_{p} = 130 \bigg[ \bigg(\frac{E}{10^{50}\erg}\frac{10^{-3}\cm^{-3}}{n}\bigg)^{1/3} \bigg]
 \bigg(\frac{\theta_{\rm{obs}}-\theta_j}{15^\circ}\bigg)^2 \rm{days} \ ,
\end{equation}
where $ E $ is the total jet's energy, $ \epsilon_e $  and $ \epsilon_B $ are the electron and magnetic field equipartition parameters, $ n $ is the  circum-merger number density, taken to be uniform in space, $ D $ is the distance to Earth, $ \nu $ is the observed frequency and $ g(p) \approx 10^{-0.31p}\Big(\frac{p-2}{p-1}\Big)^{p-1} $ is a normalization numerical factor. 

Eqs. \ref{eq:peak_magn} and \ref{eq:peak_timen} are normalized according to canonical values inferred for GW170817. The different terms in these equations 
are arranged according to their origin. In curly bracket we have terms that depends on the microphysical parameters that are least known. We will not vary those in the rest of the text. In square brackets we have the energy and the external density. We keep those fixed in section \ref{sec:GW17like_model} in which we discuss GW170817-like events and vary only the viewing angle and distance to the source. We vary the parameters that are in square brackets in subsequent discussion when considering events with different energies and circum-merger densities.

Eqs. \ref{eq:peak_magn} and \ref{eq:peak_timen} are applicable as long as the considered frequency is above the self absorption and below the cooling frequency, as was the case for GW170817. Using the analytic relations in \citet{Granot2002} and \citet{Wijers1999} and assuming an electron equipartition parameter $ \epsilon_e = 0.1 $, we find that self-absorption may affect the flux at $ \nu = 3\rm{GHz} $ only at relativity large densities $ n \gtrsim n_a \approx 10\cm^{-3} $. The value of $n_a$ is weakly dependent on the other system's parameters. Therefore we are able to scan a broad range of densities for which the afterglow at $3$GHz is in the same spectral regime, as the multi-band afterglow observations of GW170817.  Moreover, while  we focus here on radio observations, our results are not limited to radio bands (see $ \S $ \ref{sec:optical_xray}).

\section{Afterglow detectability  of GW events}
\label{sec:detectability}
We turn  now to use the numerically calibrated analytic expressions for the peak flux and time,
Eqs. \ref{eq:peak_magn} and \ref{eq:peak_timen}, to explore the detectability of future afterglows from double NS mergers.
We assume $ p = 2.16 $ and a  jet opening angle $\theta_j = 5^\circ$, as implied by the observations of the afterglow of GW170817 \citep{Mooley2018a}.
For the purpose of estimating the detectability we {use}  the current detection limit of Karl G. Jansky Very Large Array (JVLA) to be $ F_{\rm lim} \approx 10\mu \rm{Jy} $ at $ \nu = 3\rm{GHz} $. {The detectability can} be scaled trivially for different limiting fluxes.

{We begin by considering GW170817-like events. These are events with  the same parameters as those inferred from the afterglow of GW170817. The considerations here are almost model-independent as they are independent of all the specific parameters we have chosen when modeling the afterglow of GW170817, except for the viewing angle.
Later on we consider events with other circum-merger densities and jet's energies, as these parameters are expected to vary from one merger to another.
We do not vary the microphysical parameters. These parameters depend on the particle acceleration and magnetic field amplification mechanisms at the shock. They depend only on the local conditions at the shock and are independent of global parameters such as the jet energy and opening angle or the circum-merger density. As the blast wave producing the afterglow at $\theta_{obs} \gg \theta_j$ is expected to be in the same physical regime for all future mergers (i.e., similar Lorentz factors), it is reasonable that {these parameters do not change significantly} between different events.
Finally, we emphasize that while we present our estimates in the radio band, as long as the radio, optical and X-rays are in the same spectral range (above the  self absorption and the typical synchrotron frequencies and below the cooling frequency), the results apply for optical and X-ray bands as well 
(see  $ \S $ \ref{sec:optical_xray}). }

\subsection{GW170817-like events}
\label{sec:GW17like_model}

We examine, first, the detection horizon and the detection rate of GW170817-like events.
These events are defined as having the same jet's energy, circum-merger density and micro-physical parameters as GW170817.
For such events, we normalize Eqs. \ref{eq:peak_magn} and \ref{eq:peak_timen} using observed values in GW170817, 
$F_{\nu,p,\rm GW17} = 90 \mu {\rm Jy}$ at 3GHz and $t_{p,\rm GW17}=130$ days, and keeping only the dependence on the viewing and jet angles ($ \theta_{\rm obs}$ and $\theta_j$) and the distance to earth $ D $. This enables us to avoid the dependence on the less certain parameters, marked in curly and square brackets in Eq. \ref{eq:peak_magn}, making the analysis for GW170817-like events very robust. 
The remaining scaling values in these equations are all obtained from the observations of GW170817. The electrons' power-law index was found to be $ p = 2.16 $ \citep{Mooley2018a,Alexander2018,Margutti2018}, and the distance was $ D_{\rm GW17} \approx 41_{-3}^{+3}\rm Mpc $ \citep{Hjorth2017,Cantiello2018,Lee2018}. \citet{Mooley2018b} estimated the viewing angle of GW170817, to be  $ \theta_{\rm obs,GW17} = 20^\circ $, within the range   $14^\circ < \theta_{\rm{obs,GW17}} <29^\circ $,
in agreement with estimates based on the gravitational radiation signal \citep{Abbott2017,Cantiello2018}, and the jet opening angle $\theta_{j,\rm GW17}=5^\circ$.

The viewing angle that we use is scaled relative to the viewing angle of  GW170817, $ \theta_{\rm{obs,GW17}}$. The range of values of   $ \theta_{\rm{obs,GW17}}$ is the main source of uncertainty in the estimate of the detectability of GW17081-like events, leading to a variation by a factor of 5 in the flux estimate and a factor of 2.5 in the peak time estimate.  

\begin{figure}
	\centering
	\includegraphics[width=0.5\textwidth]{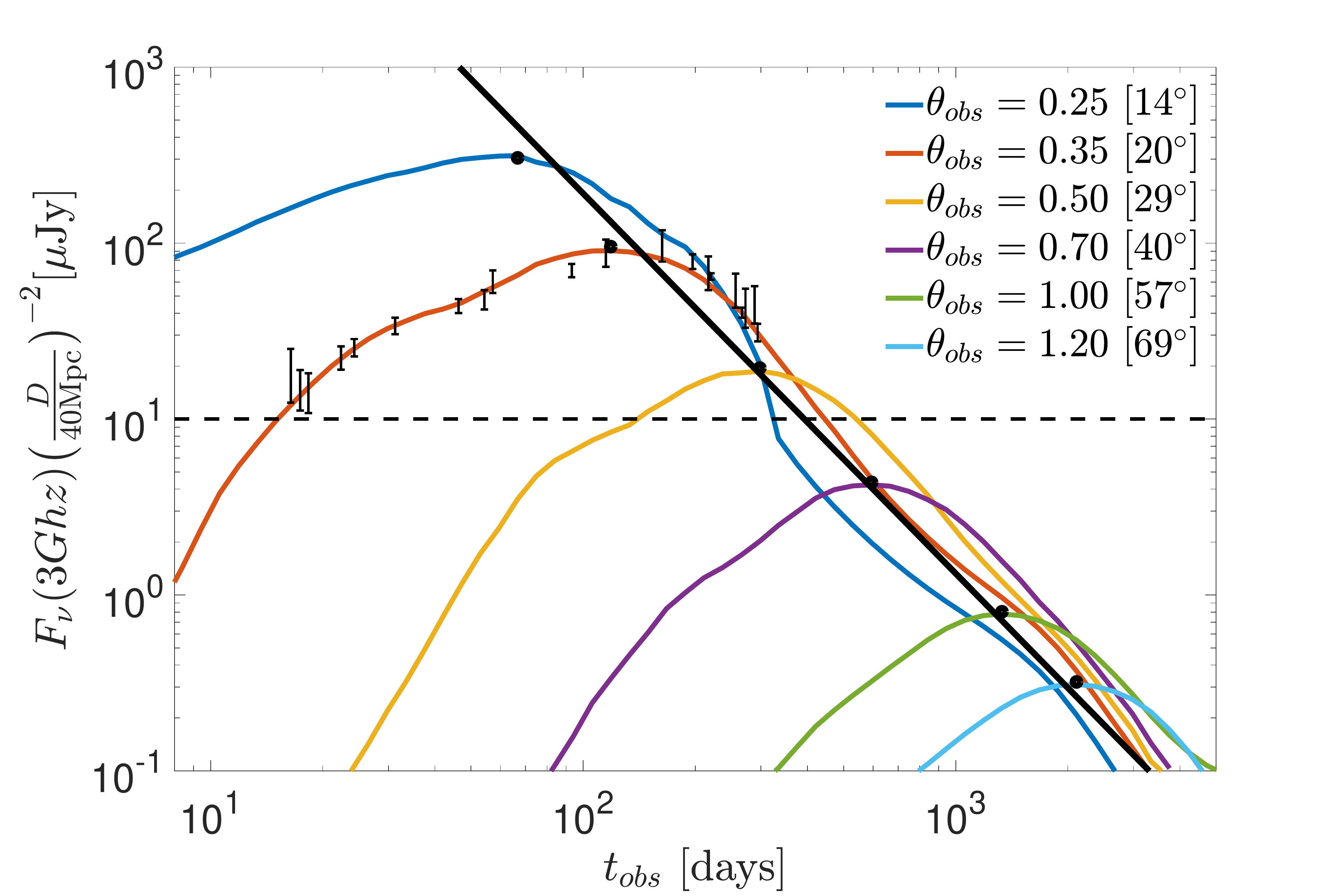}
	\caption[Light curves for different observers]{
		The $ \nu = 3\rm{GHz} $ numerical light curves (simulation $ \A $) at different viewing angles.
		The analytic relation $ F_{\nu,p} \propto t^{-p} $ ({black curve}) between the time of the peak (black dots) and the viewing angle fits nicely the numerical simulations. The horizontal dashed black line is the detection limit, approximated here as $ F_{\rm lim} = 10\mu \rm{Jy} $.
		The error bars on the red curve depict the observed data points GW170817 from \citet{Hallinan2017,Alexander2017,Alexander2018,Dobie2018,Margutti2018,Mooley2018a,Mooley2018c}.
	}
	\label{fig:angles}
\end{figure}

Figure \ref{fig:angles} depicts a numerical light curve that fits the observed GW170817 afterglow when observed at $ 20^\circ $ (red line). We also present the light curves that the simulation yields at different viewing angles (we consider $\theta_{\rm obs} > 14^\circ$  \footnote{Calculations of observers closer to the axis would require somewhat different numerical schemes than the one we use.}). The horizontal dashed black line reflects the detection threshold of JVLA.  This indicates at which angles GW170817 could have been detected.
For the canonical viewing angle  $ \theta_{\rm{obs,GW17}}= 20^\circ$, GW170817 was detectable up to $ \theta_{\rm{obs}} = \frac{5}{3}\theta_{\rm{GW17}} \approx 33^\circ $, namely we can detect the radio afterglow of about one out of six\footnote{Assuming, of course, that the events are uniformly distributed in viewing angle.}
 GW170817-like events at the distance  $ D_{\rm GW17} = 40 \rm Mpc$.
The only  uncertainty in this estimate is the uncertainty in, $ \theta_{\rm{obs,GW17}}$. Using the upper (lower) limit of  
$\theta_{\rm{obs,GW17}} = 29^\circ (14^\circ) $, then  a third (1/12) of GW170817-like events at $ D_{\rm GW17} $ can be detected with current radio telescopes.

\begin{figure}
	\centering
	\includegraphics[width=0.5\textwidth]{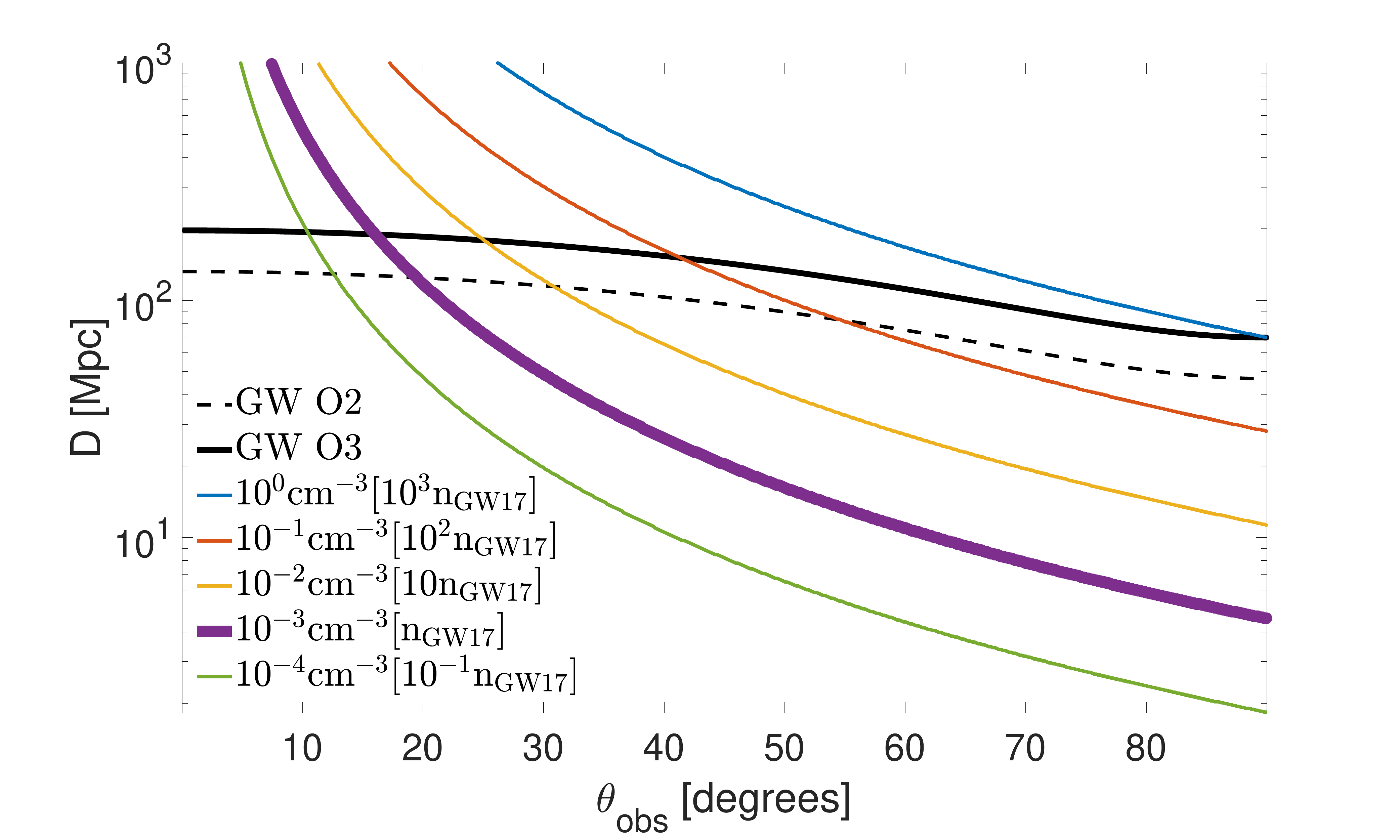}
	\caption[Probability]{
		The radio detection horizon (for $F_{\rm lim} = 10\mu \rm{Jy} $) of GW170817-like afterglow, for the energy of GW170817 and a variety of circum-merger densities (color lines).  The thick purple line depicts the canonical circum-merger density of GW170817. Also shown are the detection horizons of Advanced LIGO/Virgo observation runs O2 (black dashed line) and O3 (black solid line).
	}
	\label{fig:probability}
\end{figure}

In Figure \ref{fig:probability} we show the detection horizon of a GW170817-like  afterglow (purple thick line) at different viewing angles. Also shown are the GW170817 gravitational waves detection horizons in the Advanced LIGO/Virgo observation run O2 and in the upcoming run O3 (in black). The dependence of gravitational waves detection horizon upon the viewing angle is approximated by:
$ D_{\rm{GW}}(\theta) = D_0 \Big[\big(1+6\rm{cos}^2\theta+\rm{cos}^4\theta\big)/8\Big]^{1/2} $ \citep{Schutz2011},
where $ D_0 = 132~(197) \rm{Mpc} $ is the detection horizon of a face-on neutron star binary system in O2 (O3)  \citep{Chen2017}.
The detection horizon of the afterglow at small angles is much larger than that of gravitational waves, and vice versa at large angles. At $ \theta_{\rm{obs}} = 16^\circ $ the GW detection horizon of run O3, 190Mpc, equals to that of a GW170817-like afterglow.
The detection horizon of the afterglow decreases rapidly with the viewing angle. At $ \theta_{\rm{obs}} = 30^\circ $ it is only 50Mpc.

We can use Eqs. \ref{eq:peak_magn} and \ref{eq:peak_timen} scaled to the values of GW170817 to calculate the prevalence of joint detections of GW and afterglow from GW170817-like events. Namely, in what percentage of GW detections, we can also  detect an afterglow counterpart.
In run O3, only $ 18\% $ of all GW detections from binary NS mergers should be accompanied by a detectable GW170817-like afterglow, typically at small viewing angles.
At $ \theta_{\rm{obs}} \lesssim 16^\circ $ all GW detections of GW170817-like events in O3 should be accompanied by an detectable afterglow.
However, due to the strong sensitivity of the afterglow horizon on the viewing angle, already at $ \theta_{\rm{obs}} = 22^\circ $ only $ 15\% $ of GW detections will have a detectable GW170817-like afterglow counterpart.
It is therefore most likely to have a joint detection around $ \theta_{\rm{obs}} = 15^\circ-20^\circ $, similar to what we have seen in GW170817.

\subsection{Afterglows with various jet energies and circum-merger densities}
\label{sec:varying_parameters}

The next binary neutron star (BNS)  merger event detectable in GW  will most likely take place in a different environment than GW170817 and the energy of its jet may also be different. 
Keeping the microphysics parameters fixed, we study next the detectability of jets with various energies and densities.
We use Eqs. \ref{eq:peak_magn} and \ref{eq:peak_timen} to predict the detectability of an event with  a given energy, circum-merger  density and viewing angle. Now we vary the terms in square brackets in Eq. \ref{eq:peak_magn} as well. 
However,  to do so we need first to estimate $ E_{\rm GW17} $ and $ n_{\rm GW17} $. We use $E_{\rm GW17} = 10^{50\pm 1 }\erg $, $ n_{\rm GW17} =10^{-3^{+0.3}_{-1}} \cm^{-3} $  and $ \theta_{\rm obs,GW17} = 20^{\circ +{9^{\circ}}}_{-6^{\circ} }  $ \citep[please see][for a detailed explanation of the derivation of these values]{Mooley2018b} .
Like in the previous discussion  (see \S \ref{sec:GW17like_model})  the main source of uncertainty in the detectability predictions is the uncertainty in these estimates of the exact values of the afterglow of GW170817.

We consider mergers with a wide range of energies and densities. When varying the circum-merger density, we consider densities that are below the critical density $ n_a \approx 10\cm^{-3} $ so that self-absorption does not play a role here and the analytic relations are valid at $ \nu = 3\rm GHz $.
Figure \ref{fig:probability} depicts the detection horizon for a variety of densities, assuming $ E = E_{\rm GW17} $. At the higher density shown,  $ n \approx 1\cm^{-3} $, all GW detections from NS mergers in run O3 are expected to be accompanied by a detectable afterglow.

The majority of GW detections is expected to be at $ \theta_{\rm obs} = 30^\circ $ \citep{Schutz2011}. Therefore, at densities for which the afterglow detection horizon is larger than that of GW at $ \theta_{\rm obs} = 30^\circ $, we expect to detect a large fraction of the  events. In particular, at $ n \gtrsim 5\times 10^{-2}\cm^{-3} $ which seems to be rather common in sGRBs \citep{Berger2014}, more than $ 70\% $ of afterglows with $E \approx  E_{\rm GW17}$ that accompany a GW signal can be detected.
At lower densities than $ n_{\rm GW17} $, only a few percent of GW from NS binaries in run O3 will provide a detectable afterglow counterpart, with a small typical viewing angle of $ \theta_{\rm obs} \approx 10^\circ $.
Therefore, most GW-EM detections are expected to take place at regions with densities of $ \sim 0.1\cm^{-3} $  where the afterglow signal also appears at rather early times and is detectable at $ \theta_{\rm obs} \approx 30^\circ $.

\begin{figure}
	\centering
	\includegraphics[width=0.5\textwidth]{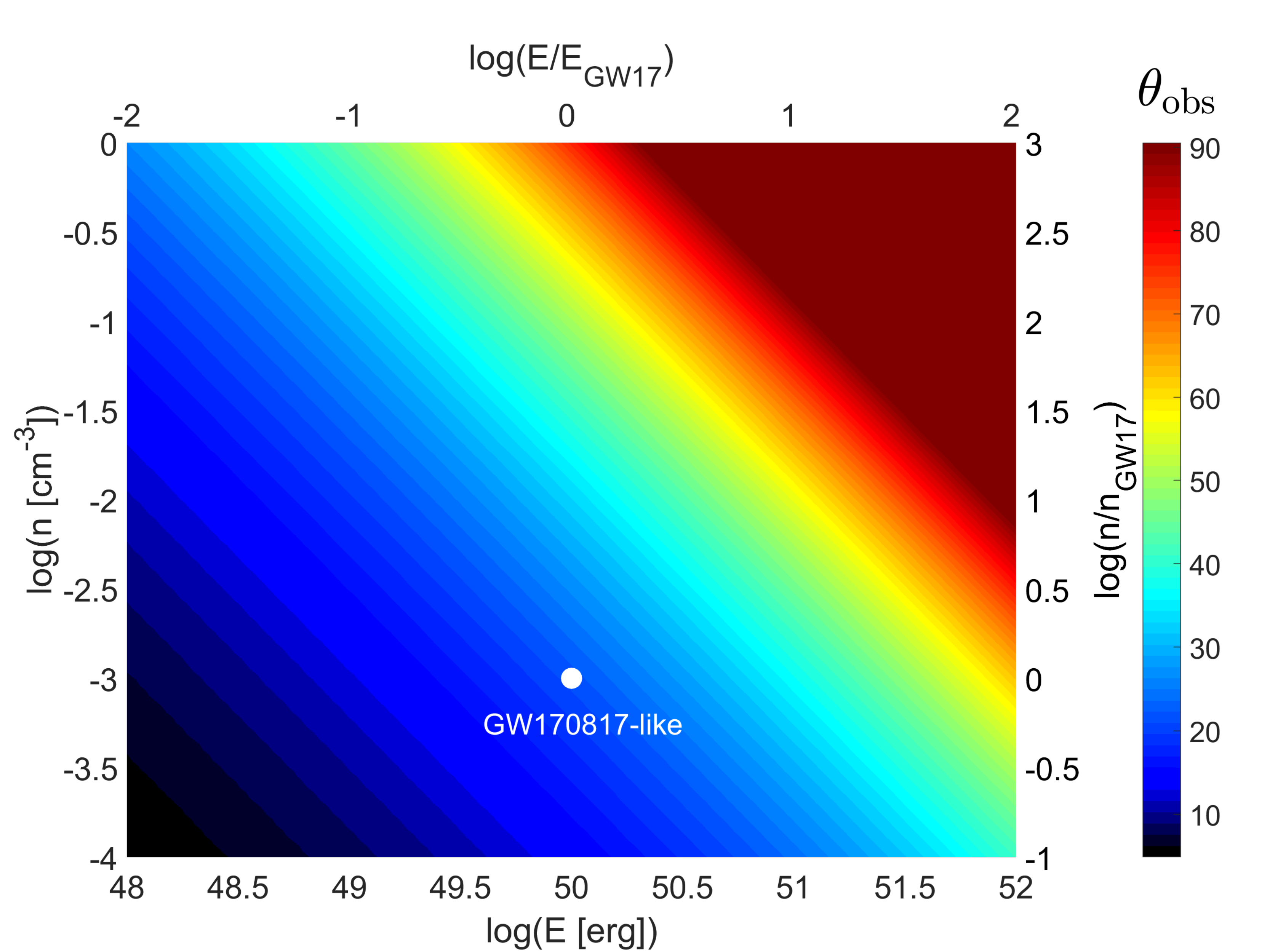}
	\caption[Light curves for different densities]{
		The maximal detectable $ \theta_{\rm obs} $ (color scale in degrees) for different circum-merger densities and jet's energies at $ D = 100\rm Mpc $.
		In the right and top axes we depict the densities and energies in respect to those in GW170817. In the left and bottom axes we use the estimates of GW170817 parameters to present the densities and energies in absolute values.
		We assume the detection limit to be $ F_{\rm lim} = 10\mu \rm Jy $.
 The white dot represents the best fit to GW170817.
	}
	\label{fig:peaks_map}
\end{figure}

The observed GW signal provides the distance (as well as a constraint on the viewing angle) of the merger. This allows us to consider the effect of the circum-merger density and the jet's energy on the maximal viewing angle at which an afterglow can be detected.
Figure \ref{fig:peaks_map} depicts this angle at a distance of $ D = 100\rm Mpc $ for a variety of $ n $ and $ E $, assuming $ F_{\rm lim} = 10\mu \rm{Jy} $. A GW170817-like event is detectable at this distance up to $ \theta_{\rm{obs}} \approx 20^\circ $, namely can be detected with a probability of $ 7\% $. Since the detectability scales with the circum-merger density, energy, distance and the detection limit, we can approximate for small angles the maximal viewing angle with a detectable afterglow.
Inverting Eq. \ref{eq:peak_magn} we obtain:
\small
\begin{equation}
\theta_{\rm obs,max} = 20^\circ \Big(\frac{n}{ 10^{-3}\cm^{-3}}\Big)^{0.18}\Big(\frac{D}{100\rm{Mpc}}\Big)^{-0.46}\Big(\frac{E}{10^{50}\erg}\frac{10\mu\rm{Jy}}{F_{\rm lim}}\Big)^{0.23} \ .
\end{equation}\normalsize

\subsection{Maximal Time for Detection}
\label{sec:latest_times}

An interesting observational question is what is the latest time we can expect a detection of  an afterglow. Namely, supposed that we have not detected an afterglow for some period after the merger, should we keep on searching?  Since detection is easiest at the peak, this translates to what is the latest time, $t_{\rm last}$, we can detect {the peak of} a given event.
Large circum-merger densities give rise to a strong early peak whereas low densities yield faint afterglows that peak at late time. For each distance, energy and viewing angle we can find the minimal density for which $F_{\nu,p} > F_{\rm lim} $. 
This density corresponds to the latest time that an afterglow can be detected. Namely, if by this time there was no detection, the afterglow from the jet of this event will never be detected.
We use Eqs. \ref{eq:peak_magn} and \ref{eq:peak_timen} to obtain\footnote{We ignore here the  weak dependence on the viewing angle and take a conservative value for $\theta_{\rm obs} = 90^\circ$.}:
\begin{equation}\label{eq:latest_time}
{         
	\begin{split}
	t_{\rm{last}} \approx 230 ~{\rm days} \Big(\frac{E}{10^{50}\erg}\Big)^{0.76}\Big(\frac{\epsilon_B}{6\times 10^{-5}}\Big)^{1/3}\Big(\frac{\epsilon_e}{0.1}\Big)^{0.49} \times\\\Big(\frac{F_{\rm lim}}{10\mu\rm{Jy}}\Big)^{-0.42}\Big(\frac{D}{150\rm{Mpc}}\Big)^{-0.84} \ .
	\end{split}}
\end{equation}
For events with jet energies similar to GW170817, we find that at a typical distance for a detection during O3, $ D \approx 150\rm Mpc $, the latest time for detection is $ \sim 230 ~\rm days $.
For a very energetic jet, $ E \approx  10^{51}{\rm erg} ~\approx 10 E_{\rm GW17} $, or alternatively an event with a higher magnetic equipartition parameter, $ \epsilon_B = 0.1 $, the afterglow can be detected up to $ \sim 2 $ years after the merger.
Eq. \ref{eq:latest_time} implies that if we have not detected a jet signal up to some time, then there is a lower limit on the jet energy needed to enable a detection at later times. For example, an event at a distance $ D  = 200\rm{Mpc} $ which has not shown an afterglow signal during the first 100 days after the merger, must involve a jet with energy $ E \gtrsim 10^{50}\erg $ to have a chance of being detectable in the future. This result is only weakly dependent on the viewing angle.

Finally, we stress that Eq. \ref{eq:latest_time} considers the latest time for a detection of an afterglow from a  relativistic jet.
The merger is expected to have also a mildly to sub-relativistic ejecta which may contain much more energy than the relativistic jet. Moreover, the microphysical parameters of non-relativistic shocks, such as $\epsilon_B$, may be different than those of relativistic shocks. Therefore, even if there is no detection of a relativistic jet, other ejecta components may still be detectable in the radio on longer timescales \citep{Nakar2011}.

\subsection{Optical and  X-ray}
\label{sec:optical_xray}

The electromagnetic follow-up takes place also in the IR, optical and X-ray bands. The optical and IR bands are not affected by either cooling or self absorption, and hence Eqs. \ref{eq:peak_magn} and \ref{eq:peak_timen} are applicable to the optical and IR bands for any reasonable values of the circum-merger density and microphysical parameters.
The detection limit we estimated for the radio, $ F_{\rm lim} = 10\mu \rm Jy $, is equivalent to a magnitude of $ \sim 29.2 $ for $ p = 2.16 $. This value is above the HST limit in the GW170817 observations \citep{Lyman2018,Margutti2018}, and therefore HST is not expected to detect future afterglows that are not detectable in the radio bands. Furthermore, unlike the radio, observations in the optical bands are constrained by the location of the Sun.
Therefore, the optical afterglow does not affect the detectability results that we derived\footnote{ This discussion applies only to the afterglow, the macronova signal has very different detectability considerations. }.

If the X-ray afterglow is on the same spectral regime as the other afterglow bands, as observed in GW170817,
then the detectability of the afterglow by Chandra X-ray observatory and XMM-Newton is similar to its detectability by the JVLA\footnote{An exact  comparison between the detection limit of the JVLA and Chandra X-ray observatory depends very strongly on the value of $p$.} (as was in the case of GW170817), with the caveat of the observational constraints posed by the Sun. 
However, for a sufficiently large  circum-merger density, the X-ray light curve can be affected by a cooling break. For $ h\nu = 1\rm keV$ cooling is relevant   for  $ n \ge 69 ({\epsilon_B}/{6\times 10^{-5}})^{-3/2} [({t}/{130\rm d} )({E}/{10^{50}\erg})]^{-1/2} \cm^{-3} $ \citep{Wijers1999,Granot2002}.
While for our best fit GW170817 parameters, the density for which cooling is important is quite large, different parameters can induce cooling at lower densities. In particular, the strong dependence on the poorly constrained magnetic equipartition parameter $ \epsilon_B $, may place the critical density much lower than $ n = 1\cm^{-3} $. For such cases, our results in $ \S $ \ref{sec:varying_parameters} do not apply to the X-ray bands.

\section{Conclusions}
\label{sec:conclusions}

The gravitational waves signal of the double neutron star merger, GW170817, were followed by a series of electromagnetic signals: a weak sGRB, a macronova and a multi-wavelengths afterglow. 
  \citet{Mooley2018b} used VLBI observations of the afterglow and showed that at our viewing angle, $ \theta_{\rm obs} = 20^\circ $, the afterglow was dominated by the cocoon at first, and then the  jet's core dominated at the peak and the subsequent decline. In this work we explored the detectability of future afterglow of merger events in view of what we have learnt from GW170817. We considered viewing angles that are larger than the jet opening angle ($ \theta_{\rm obs} \gg \theta_j $) and assume that  similar to GW170817, the afterglow powered by the jet dominates at the peak. We have shown that in this case the afterglow of the merger at the peak resembles the one expected from the  afterglow of an equivalent top-hat jet seen off-axis.

We first demonstrated using a set of numerical simulations that the analytic equations in \citet{Nakar2002} which describe the orphan afterglow of top-hat jets are generic and apply to all types of GRBs. Including those with complicated angular structure surrounding the jet, as long as the jet's core contains most of the energy.
Then,  using these numerical simulations we calibrated the numerical factor in the analytic equations of \citet{Nakar2002} and  derive expressions for the peak flux and time for a given set of parameters.
As the peak of the afterglow determines the detectability of the event, we used these expressions to derive the detection horizon and detection rates of future events.

We  considered the detectability of GW170817-like systems as a function of the distance and viewing angle. These are systems {whose afterglow is similar to that} of GW170817 (that is with a similar geometry, energy, circum-merger density and microphysical parameters). 
We showed that a joint detection of GW and afterglow from GW170817-like systems is rare, with merely $ 7\% $ of GW170817-like afterglows are detectable at $ D = 100\rm Mpc $. In the upcoming Advanced LIGO/VIRGO run O3, GW170817-like afterglows will be detectable only at small angles ($ \theta_{\rm{obs}} \lesssim 20^\circ $), and thus only for about $ \sim 20\% $ of GW170817-like  binary NS mergers.
Thus, if GW170817 represents a typical NS merger, future joint detections of both GW and afterglows may not be very common.
The main uncertainty in these estimates arises from the uncertainty of the viewing angle of GW170817. Taking this into account will amount to an uncertainty of a factor of two below or above the aforementioned rates.

A higher circum-merger density results in an earlier and stronger peak. Our calculations show that for typical circum-merger densities $ n \gtrsim 5\times 10^{-2} \cm^{-3} $, GW-EM detections will be more common ($ \gtrsim 70\% $), most of which at $ \theta_{\rm{obs}} \approx 30^\circ $.
It is interesting that we can also estimate the latest time after which we can conclude that the jet afterglow counterpart is not detectable.
We find that the maximal peak time (after which detection is not expected) is largely independent of the viewing angle (see Eq. \ref{eq:latest_time} and subsequent discussion), and it is mainly governed by the distance and the jet's energy.
At a typical distance of $ D \approx 150 \rm{Mpc} $, had an afterglow with an energy similar to that in GW170817, not been detected after $ \sim 230\rm{days} $, it cannot be detected later.

Finally, our analysis applies for events that their afterglow peak is dominated by the jet's core. It is plausible that at least some of the afterglows that accompany GW from binary NS will show very little in common with GW170817. One particular example is the case of a choked jet. Arguments which suggested that the majority of jets from collapsing stars never break out \citep{Mazzali2008,Bromberg2011,Sobacchi2017}, may still hold in sGRBs as well. Furthermore, if the sGRB duration distribution shares similarities with long GRBs \citep{Moharana2017}, most sGRBs jets may also be choked.
If the jet is choked, the outflow expands to wide angles \citep{Gottlieb2018b,Nakar2018} with a weak dependence of the flux upon the viewing angles inside the cocoon opening angle, which may be as large as $ \theta \approx 60^\circ $. The peak flux of a choked jet afterglow depends, among other things, on the total energy deposited into the cocoon and on the jet opening angle, and at large viewing angles it may be higher or lower than that of GW170817.
Note that choked jets may produce a strong $ \gamma $-ray emission via cocoon shock-breakout at large viewing angles \citep{Gottlieb2018b}, allowing a joint GW/$ \gamma $-ray/afterglow detection to be more common.

\section*{Acknowledgements}
This research is supported by the CHE-ISF I-Core center for excellence in Astrophysics. 
 TP is partially supported by an advanced ERC grant TReX. OG and EN are partially supported by a consolidator ERC grant (JetNS) and an ISF grant.

\bibliographystyle{mnras} 
\bibliography{angles}

\begin{thebibliography}{}
\makeatletter
\relax
\def\mn@urlcharsother{\let\do\@makeother \do\$\do\&\do\#\do\^\do\_\do\%\do\~}
\def\mn@doi{\begingroup\mn@urlcharsother \@ifnextchar [ {\mn@doi@}
  {\mn@doi@[]}}
\def\mn@doi@[#1]#2{\def\@tempa{#1}\ifx\@tempa\@empty \href
  {http://dx.doi.org/#2} {doi:#2}\else \href {http://dx.doi.org/#2} {#1}\fi
  \endgroup}
\def\mn@eprint#1#2{\mn@eprint@#1:#2::\@nil}
\def\mn@eprint@arXiv#1{\href {http://arxiv.org/abs/#1} {{\tt arXiv:#1}}}
\def\mn@eprint@dblp#1{\href {http://dblp.uni-trier.de/rec/bibtex/#1.xml}
  {dblp:#1}}
\def\mn@eprint@#1:#2:#3:#4\@nil{\def\@tempa {#1}\def\@tempb {#2}\def\@tempc
  {#3}\ifx \@tempc \@empty \let \@tempc \@tempb \let \@tempb \@tempa \fi \ifx
  \@tempb \@empty \def\@tempb {arXiv}\fi \@ifundefined
  {mn@eprint@\@tempb}{\@tempb:\@tempc}{\expandafter \expandafter \csname
  mn@eprint@\@tempb\endcsname \expandafter{\@tempc}}}

\bibitem[\protect\citeauthoryear{Abbott et~al.,}{Abbott
  et~al.}{2017}]{Abbott2017}
Abbott B.~P.,  et~al., 2017, \mn@doi [Nature, Volume 551, Issue 7678, pp. 85-88
  (2017).] {10.1038/nature24471}, 551, 85

\bibitem[\protect\citeauthoryear{Alexander et~al.,}{Alexander
  et~al.}{2017}]{Alexander2017}
Alexander K.~D.,  et~al., 2017, \mn@doi [The Astrophysical Journal Letters,
  Volume 848, Issue 2, article id. L21, 7 pp. (2017).]
  {10.3847/2041-8213/aa905d}, 848

\bibitem[\protect\citeauthoryear{Alexander et~al.,}{Alexander
  et~al.}{2018}]{Alexander2018}
Alexander K.~D.,  et~al., 2018, \mn@doi [The Astrophysical Journal Letters,
  Volume 863, Issue 2, article id. L18, 6 pp. (2018).]
  {10.3847/2041-8213/aad637}, 863

\bibitem[\protect\citeauthoryear{Andreoni et~al.,}{Andreoni
  et~al.}{2017}]{Andreoni2017}
Andreoni I.,  et~al., 2017, \mn@doi [Publications of the Astronomical Society
  of Australia, Volume 34, id.e069 21 pp.] {10.1017/pasa.2017.65}, 34

\bibitem[\protect\citeauthoryear{Arcavi et~al.,}{Arcavi
  et~al.}{2017a}]{Arcavi2017}
Arcavi I.,  et~al., 2017a, \mn@doi [Nature, Volume 551, Issue 7678, pp. 64-66
  (2017).] {10.1038/nature24291}, 551, 64

\bibitem[\protect\citeauthoryear{Arcavi et~al.,}{Arcavi
  et~al.}{2017b}]{Arcavi2017a}
Arcavi I.,  et~al., 2017b, \mn@doi [The Astrophysical Journal Letters, Volume
  848, Issue 2, article id. L33, 12 pp. (2017).] {10.3847/2041-8213/aa910f},
  848

\bibitem[\protect\citeauthoryear{Beloborodov, Lundman  \& Levin}{Beloborodov
  et~al.}{2018}]{Beloborodov2018}
Beloborodov A.~M.,  Lundman C.,   Levin Y.,  2018, eprint arXiv:1812.11247

\bibitem[\protect\citeauthoryear{{Beniamini} \& {Nakar}}{{Beniamini} \&
  {Nakar}}{2019}]{Beniamini2019a}
{Beniamini} P.,  {Nakar} E.,  2019, \mn@doi [\mnras] {10.1093/mnras/sty3110},
  \href {http://cdsads.u-strasbg.fr/abs/2019MNRAS.482.5430B} {482, 5430}

\bibitem[\protect\citeauthoryear{{Beniamini}, {Petropoulou}, {Barniol Duran}
  \& {Giannios}}{{Beniamini} et~al.}{2019}]{Beniamini2019b}
{Beniamini} P.,  {Petropoulou} M.,  {Barniol Duran} R.,   {Giannios} D.,  2019,
  \mn@doi [\mnras] {10.1093/mnras/sty3093}, \href
  {http://cdsads.u-strasbg.fr/abs/2019MNRAS.483..840B} {483, 840}

\bibitem[\protect\citeauthoryear{Berger}{Berger}{2014}]{Berger2014}
Berger E.,  2014, \mn@doi [Annual Review of Astronomy and Astrophysics, vol.
  52, p.43-105] {10.1146/annurev-astro-081913-035926}, 52, 43

\bibitem[\protect\citeauthoryear{Bromberg, Nakar  \& Piran}{Bromberg
  et~al.}{2011}]{Bromberg2011}
Bromberg O.,  Nakar E.,   Piran T.,  2011, \mn@doi [Astrophysical Journal
  Letters] {10.1088/2041-8205/739/2/L55}, 739

\bibitem[\protect\citeauthoryear{Bromberg, Tchekhovskoy, Gottlieb, Nakar  \&
  Piran}{Bromberg et~al.}{2017}]{Bromberg2017}
Bromberg O.,  Tchekhovskoy A.,  Gottlieb O.,  Nakar E.,   Piran T.,  2017,
  \mn@doi [Monthly Notices of the Royal Astronomical Society, Volume 475, Issue
  3, p.2971-2977] {10.1093/mnras/stx3316}, 475, 2971

\bibitem[\protect\citeauthoryear{Buckley et~al.,}{Buckley
  et~al.}{2017}]{Buckley2017}
Buckley D. A.~H.,  et~al., 2017, \mn@doi [Monthly Notices of the Royal
  Astronomical Society: Letters, Volume 474, Issue 1, p.L71-L75]
  {10.1093/mnrasl/slx196}, 474, L71

\bibitem[\protect\citeauthoryear{Cantiello et~al.,}{Cantiello
  et~al.}{2018}]{Cantiello2018}
Cantiello M.,  et~al., 2018, \mn@doi [The Astrophysical Journal]
  {10.3847/2041-8213/aaad64}, 854, L31

\bibitem[\protect\citeauthoryear{Chen, Holz, Miller, Evans, Vitale  \&
  Creighton}{Chen et~al.}{2017}]{Chen2017}
Chen H.-Y.,  Holz D.~E.,  Miller J.,  Evans M.,  Vitale S.,   Creighton J.,
  2017, eprint arXiv:1709.08079

\bibitem[\protect\citeauthoryear{Chornock et~al.,}{Chornock
  et~al.}{2017}]{Chornock2017}
Chornock R.,  et~al., 2017, \mn@doi [The Astrophysical Journal Letters, Volume
  848, Issue 2, article id. L19, 7 pp. (2017).] {10.3847/2041-8213/aa905c}, 848

\bibitem[\protect\citeauthoryear{Coulter et~al.,}{Coulter
  et~al.}{2017}]{Coulter2017}
Coulter D.~A.,  et~al., 2017, \mn@doi [Science, Volume 358, Issue 6370, pp.
  1556-1558 (2017).] {10.1126/science.aap9811}, 358, 1556

\bibitem[\protect\citeauthoryear{Cowperthwaite et~al.,}{Cowperthwaite
  et~al.}{2017}]{Cowperthwaite2017}
Cowperthwaite P.~S.,  et~al., 2017, \mn@doi [The Astrophysical Journal Letters,
  Volume 848, Issue 2, article id. L17, 10 pp. (2017).]
  {10.3847/2041-8213/aa8fc7}, 848

\bibitem[\protect\citeauthoryear{D'Avanzo et~al.,}{D'Avanzo
  et~al.}{2018}]{DAvanzo2018}
D'Avanzo P.,  et~al., 2018, \mn@doi [Astronomy {\&} Astrophysics, Volume 613,
  id.L1, 5 pp.] {10.1051/0004-6361/201832664}, 613

\bibitem[\protect\citeauthoryear{D{\'{i}}az et~al.,}{D{\'{i}}az
  et~al.}{2017}]{Diaz2017}
D{\'{i}}az M.~C.,  et~al., 2017, \mn@doi [The Astrophysical Journal Letters,
  Volume 848, Issue 2, article id. L29, 5 pp. (2017).]
  {10.3847/2041-8213/aa9060}, 848

\bibitem[\protect\citeauthoryear{Dobie et~al.,}{Dobie et~al.}{2018}]{Dobie2018}
Dobie D.,  et~al., 2018, \mn@doi [The Astrophysical Journal Letters, Volume
  858, Issue 2, article id. L15, 5 pp. (2018).] {10.3847/2041-8213/aac105}, 858

\bibitem[\protect\citeauthoryear{{Dobie}, {Murphy}, {Kaplan}, {Ghosh},
  {Bannister}  \& {Hunstead}}{{Dobie} et~al.}{2019}]{Dobie2019}
{Dobie} D.,  {Murphy} T.,  {Kaplan} D.~L.,  {Ghosh} S.,  {Bannister} K.~W.,
  {Hunstead} R.~W.,  2019, arXiv e-prints, \href
  {http://adsabs.harvard.edu/abs/2019arXiv190301481D} {}

\bibitem[\protect\citeauthoryear{Drout et~al.,}{Drout et~al.}{2017}]{Drout2017}
Drout M.~R.,  et~al., 2017, \mn@doi [Science, Volume 358, Issue 6370, pp.
  1570-1574 (2017).] {10.1126/science.aaq0049}, 358, 1570

\bibitem[\protect\citeauthoryear{Evans et~al.,}{Evans et~al.}{2017}]{Evans2017}
Evans P.~A.,  et~al., 2017, \mn@doi [Science, Volume 358, Issue 6370, pp.
  1565-1570 (2017).] {10.1126/science.aap9580}, 358, 1565

\bibitem[\protect\citeauthoryear{{Ghirlanda} et~al.,}{{Ghirlanda}
  et~al.}{2019}]{Ghirlanda2019}
{Ghirlanda} G.,  et~al., 2019, \mn@doi [Science] {10.1126/science.aau8815},
  \href {https://ui.adsabs.harvard.edu/abs/2019Sci...363..968G} {363, 968}

\bibitem[\protect\citeauthoryear{Goldstein et~al.,}{Goldstein
  et~al.}{2017}]{Goldstein2017}
Goldstein A.,  et~al., 2017, \mn@doi [The Astrophysical Journal Letters, Volume
  848, Issue 2, article id. L14, 14 pp. (2017).] {10.3847/2041-8213/aa8f41},
  848

\bibitem[\protect\citeauthoryear{Gottlieb, Nakar, Piran  \&
  Hotokezaka}{Gottlieb et~al.}{2018}]{Gottlieb2018b}
Gottlieb O.,  Nakar E.,  Piran T.,   Hotokezaka K.,  2018, \mn@doi [Monthly
  Notices of the Royal Astronomical Society, Volume 479, Issue 1, p.588-600]
  {10.1093/mnras/sty1462}, 479, 588

\bibitem[\protect\citeauthoryear{Granot}{Granot}{2012}]{Granot2012}
Granot J.,  2012, \mn@doi [Monthly Notices of the Royal Astronomical Society]
  {10.1111/j.1365-2966.2012.20489.x}, 421, 2610

\bibitem[\protect\citeauthoryear{Granot, Panaitescu, Kumar  \& Woosley}{Granot
  et~al.}{2002}]{Granot2002}
Granot J.,  Panaitescu A.,  Kumar P.,   Woosley S.~E.,  2002, \mn@doi [The
  Astrophysical Journal, Volume 570, Issue 2, pp. L61-L64.] {10.1086/340991},
  570, L61

\bibitem[\protect\citeauthoryear{Granot, Gill, Guetta  \& {De Colle}}{Granot
  et~al.}{2018}]{Granot2018}
Granot J.,  Gill R.,  Guetta D.,   {De Colle} F.,  2018, \mn@doi [Monthly
  Notices of the Royal Astronomical Society, Volume 481, Issue 2, p.1597-1608]
  {10.1093/mnras/sty2308}, 481, 1597

\bibitem[\protect\citeauthoryear{Haggard, Nynka, Ruan, Kalogera, Cenko, Evans
  \& Kennea}{Haggard et~al.}{2017}]{Haggard2017}
Haggard D.,  Nynka M.,  Ruan J.~J.,  Kalogera V.,  Cenko S.~B.,  Evans P.,
  Kennea J.~A.,  2017, \mn@doi [The Astrophysical Journal Letters, Volume 848,
  Issue 2, article id. L25, 6 pp. (2017).] {10.3847/2041-8213/aa8ede}, 848

\bibitem[\protect\citeauthoryear{Hallinan et~al.,}{Hallinan
  et~al.}{2017}]{Hallinan2017}
Hallinan G.,  et~al., 2017, \mn@doi [Science] {10.1126/science.aap9855}, 358,
  1579

\bibitem[\protect\citeauthoryear{Hjorth et~al.,}{Hjorth
  et~al.}{2017}]{Hjorth2017}
Hjorth J.,  et~al., 2017, \mn@doi [The Astrophysical Journal]
  {10.3847/2041-8213/aa9110}, 848, L31

\bibitem[\protect\citeauthoryear{Hu et~al.,}{Hu et~al.}{2017}]{Hu2017}
Hu L.,  et~al., 2017, \mn@doi [Science Bulletin, Vol. 62, No.21, p.1433-1438,
  2017] {10.1016/j.scib.2017.10.006}, 62, 1433

\bibitem[\protect\citeauthoryear{Kasliwal et~al.,}{Kasliwal
  et~al.}{2017}]{Kasliwal2017}
Kasliwal M.~M.,  et~al., 2017, \mn@doi [Science] {10.1126/science.aap9455},
  358, 1559

\bibitem[\protect\citeauthoryear{Kilpatrick et~al.,}{Kilpatrick
  et~al.}{2017}]{Kilpatrick2017}
Kilpatrick C.~D.,  et~al., 2017, \mn@doi [Science, Volume 358, Issue 6370, pp.
  1583-1587 (2017).] {10.1126/science.aaq0073}, 358, 1583

\bibitem[\protect\citeauthoryear{{Lamb} et~al.,}{{Lamb}
  et~al.}{2019}]{Lamb2019}
{Lamb} G.~P.,  et~al., 2019, \mn@doi [\apjl] {10.3847/2041-8213/aaf96b}, \href
  {http://adsabs.harvard.edu/abs/2019ApJ...870L..15L} {870, L15}

\bibitem[\protect\citeauthoryear{Lazzati, Perna, Morsony, Lopez-Camara,
  Cantiello, Ciolfi, Giacomazzo  \& Workman}{Lazzati
  et~al.}{2018}]{Lazzati2018}
Lazzati D.,  Perna R.,  Morsony B.~J.,  Lopez-Camara D.,  Cantiello M.,  Ciolfi
  R.,  Giacomazzo B.,   Workman J.~C.,  2018, \mn@doi [Physical Review Letters]
  {10.1103/PhysRevLett.120.241103}, 120

\bibitem[\protect\citeauthoryear{Lee, Kang  \& Im}{Lee et~al.}{2018}]{Lee2018}
Lee M.~G.,  Kang J.,   Im M.,  2018, \mn@doi [The Astrophysical Journal]
  {10.3847/2041-8213/aac2e9}, 859, L6

\bibitem[\protect\citeauthoryear{Lipunov et~al.,}{Lipunov
  et~al.}{2017}]{Lipunov2017}
Lipunov V.~M.,  et~al., 2017, \mn@doi [The Astrophysical Journal Letters,
  Volume 850, Issue 1, article id. L1, 9 pp. (2017).]
  {10.3847/2041-8213/aa92c0}, 850

\bibitem[\protect\citeauthoryear{Lyman et~al.,}{Lyman et~al.}{2018}]{Lyman2018}
Lyman J.~D.,  et~al., 2018, \mn@doi [Nature Astronomy]
  {10.1038/s41550-018-0511-3}, 2, 751

\bibitem[\protect\citeauthoryear{Margutti et~al.,}{Margutti
  et~al.}{2017}]{Margutti2017}
Margutti R.,  et~al., 2017, \mn@doi [The Astrophysical Journal Letters, Volume
  848, Issue 2, article id. L20, 7 pp. (2017).] {10.3847/2041-8213/aa9057}, 848

\bibitem[\protect\citeauthoryear{Margutti et~al.,}{Margutti
  et~al.}{2018}]{Margutti2018}
Margutti R.,  et~al., 2018, \mn@doi [The Astrophysical Journal Letters, Volume
  856, Issue 1, article id. L18, 12 pp. (2018).] {10.3847/2041-8213/aab2ad},
  856

\bibitem[\protect\citeauthoryear{Matsumoto, Nakar  \& Piran}{Matsumoto
  et~al.}{2018}]{Matsumoto2018}
Matsumoto T.,  Nakar E.,   Piran T.,  2018, \mn@doi [Monthly Notices of the
  Royal Astronomical Society, Volume 483, Issue 1, p.1247-1255]
  {10.1093/mnras/sty3200}, 483, 1247

\bibitem[\protect\citeauthoryear{Mazzali, Valenti  \& {Della Valle}}{Mazzali
  et~al.}{2008}]{Mazzali2008}
Mazzali P.~A.,  Valenti S.,   {Della Valle} M.,  2008, \mn@doi [Science, Volume
  321, Issue 5893, pp. 1185- (2008).] {10.1126/science.1158088}, 321, 1185

\bibitem[\protect\citeauthoryear{McCully et~al.,}{McCully
  et~al.}{2017}]{McCully2017}
McCully C.,  et~al., 2017, \mn@doi [The Astrophysical Journal Letters, Volume
  848, Issue 2, article id. L32, 8 pp. (2017).] {10.3847/2041-8213/aa9111}, 848

\bibitem[\protect\citeauthoryear{Mignone, Bodo, Massaglia, Matsakos, Tesileanu,
  Zanni  \& Ferrari}{Mignone et~al.}{2007}]{Mignone2007}
Mignone A.,  Bodo G.,  Massaglia S.,  Matsakos T.,  Tesileanu O.,  Zanni C.,
  Ferrari A.,  2007, \mn@doi [The Astrophysical Journal Supplement Series]
  {10.1086/513316}, 170, 228

\bibitem[\protect\citeauthoryear{Moharana \& Piran}{Moharana \&
  Piran}{2017}]{Moharana2017}
Moharana R.,  Piran T.,  2017, \mn@doi [Monthly Notices of the Royal
  Astronomical Society: Letters, Volume 472, Issue 1, p.L55-L59]
  {10.1093/mnrasl/slx131}, 472, L55

\bibitem[\protect\citeauthoryear{Mooley et~al.,}{Mooley
  et~al.}{2018a}]{Mooley2018a}
Mooley K.~P.,  et~al., 2018a, \mn@doi [Nature, Volume 554, Issue 7691, pp.
  207-210 (2018).] {10.1038/nature25452}, 554, 207

\bibitem[\protect\citeauthoryear{Mooley et~al.,}{Mooley
  et~al.}{2018b}]{Mooley2018b}
Mooley K.~P.,  et~al., 2018b, \mn@doi [Nature] {10.1038/s41586-018-0486-3},
  561, 355

\bibitem[\protect\citeauthoryear{Mooley et~al.,}{Mooley
  et~al.}{2018c}]{Mooley2018c}
Mooley K.~P.,  et~al., 2018c, \mn@doi [The Astrophysical Journal Letters,
  Volume 868, Issue 1, article id. L11, 8 pp. (2018).]
  {10.3847/2041-8213/aaeda7}, 868

\bibitem[\protect\citeauthoryear{Nakar \& Piran}{Nakar \&
  Piran}{2011}]{Nakar2011}
Nakar E.,  Piran T.,  2011, \mn@doi [Nature, Volume 478, Issue 7367, pp. 82-84
  (2011).] {10.1038/nature10365}, 478, 82

\bibitem[\protect\citeauthoryear{Nakar, Piran  \& Granot}{Nakar
  et~al.}{2002}]{Nakar2002}
Nakar E.,  Piran T.,   Granot J.,  2002, \mn@doi [The Astrophysical Journal,
  Volume 579, Issue 2, pp. 699-705.] {10.1086/342791}, 579, 699

\bibitem[\protect\citeauthoryear{Nakar, Gottlieb, Piran, Kasliwal  \&
  Hallinan}{Nakar et~al.}{2018}]{Nakar2018}
Nakar E.,  Gottlieb O.,  Piran T.,  Kasliwal M.~M.,   Hallinan G.,  2018,
  \mn@doi [The Astrophysical Journal, Volume 867, Issue 1, article id. 18, 9
  pp. (2018).] {10.3847/1538-4357/aae205}, 867

\bibitem[\protect\citeauthoryear{Nicholl et~al.,}{Nicholl
  et~al.}{2017}]{Nicholl2017}
Nicholl M.,  et~al., 2017, \mn@doi [The Astrophysical Journal Letters, Volume
  848, Issue 2, article id. L18, 8 pp. (2017).] {10.3847/2041-8213/aa9029}, 848

\bibitem[\protect\citeauthoryear{Nynka, Ruan, Haggard  \& Evans}{Nynka
  et~al.}{2018}]{Nynka2018}
Nynka M.,  Ruan J.~J.,  Haggard D.,   Evans P.~A.,  2018, \mn@doi [The
  Astrophysical Journal Letters, Volume 862, Issue 2, article id. L19, 7 pp.
  (2018).] {10.3847/2041-8213/aad32d}, 862

\bibitem[\protect\citeauthoryear{Pian et~al.,}{Pian et~al.}{2017}]{Pian2017}
Pian E.,  et~al., 2017, \mn@doi [Nature, Volume 551, Issue 7678, pp. 67-70
  (2017).] {10.1038/nature24298}, 551, 67

\bibitem[\protect\citeauthoryear{{Pozanenko} et~al.,}{{Pozanenko}
  et~al.}{2018}]{Pozanenko2018}
{Pozanenko} A.~S.,  et~al., 2018, \mn@doi [\apjl] {10.3847/2041-8213/aaa2f6},
  \href {http://adsabs.harvard.edu/abs/2018ApJ...852L..30P} {852, L30}

\bibitem[\protect\citeauthoryear{Ruan, Nynka, Haggard, Kalogera  \& Evans}{Ruan
  et~al.}{2018}]{Ruan2018}
Ruan J.~J.,  Nynka M.,  Haggard D.,  Kalogera V.,   Evans P.,  2018, \mn@doi
  [The Astrophysical Journal Letters, Volume 853, Issue 1, article id. L4, 6
  pp. (2018).] {10.3847/2041-8213/aaa4f3}, 853

\bibitem[\protect\citeauthoryear{Sari, Piran  \& Narayan}{Sari
  et~al.}{1997}]{Sari1997}
Sari R.,  Piran T.,   Narayan R.,  1997, \mn@doi [The Astrophysical Journal,
  Volume 497, Issue 1, pp. L17-L20.] {10.1086/311269}, 497, L17

\bibitem[\protect\citeauthoryear{Savchenko et~al.,}{Savchenko
  et~al.}{2017}]{Savchenko2017}
Savchenko V.,  et~al., 2017, \mn@doi [The Astrophysical Journal Letters, Volume
  848, Issue 2, article id. L15, 8 pp. (2017).] {10.3847/2041-8213/aa8f94}, 848

\bibitem[\protect\citeauthoryear{Schutz}{Schutz}{2011}]{Schutz2011}
Schutz B.~F.,  2011, \mn@doi [Classical and Quantum Gravity, Volume 28, Issue
  12, id. 125023 (2011).] {10.1088/0264-9381/28/12/125023}, 28

\bibitem[\protect\citeauthoryear{Shappee et~al.,}{Shappee
  et~al.}{2017}]{Shappee2017}
Shappee B.~J.,  et~al., 2017, \mn@doi [Science, Volume 358, Issue 6370, pp.
  1574-1578 (2017).] {10.1126/science.aaq0186}, 358, 1574

\bibitem[\protect\citeauthoryear{Smartt et~al.,}{Smartt
  et~al.}{2017}]{Smartt2017}
Smartt S.~J.,  et~al., 2017, \mn@doi [Nature, Volume 551, Issue 7678, pp. 75-79
  (2017).] {10.1038/nature24303}, 551, 75

\bibitem[\protect\citeauthoryear{Soares-Santos et~al.,}{Soares-Santos
  et~al.}{2017}]{Soares-Santos2017}
Soares-Santos M.,  et~al., 2017, \mn@doi [The Astrophysical Journal Letters,
  Volume 848, Issue 2, article id. L16, 7 pp. (2017).]
  {10.3847/2041-8213/aa9059}, 848

\bibitem[\protect\citeauthoryear{Sobacchi, Granot, Bromberg  \&
  Sormani}{Sobacchi et~al.}{2017}]{Sobacchi2017}
Sobacchi E.,  Granot J.,  Bromberg O.,   Sormani M.~C.,  2017, \mn@doi [Monthly
  Notices of the Royal Astronomical Society] {10.1093/MNRAS/STX2083}, 472, 616

\bibitem[\protect\citeauthoryear{Tanvir et~al.,}{Tanvir
  et~al.}{2017}]{Tanvir2017}
Tanvir N.~R.,  et~al., 2017, \mn@doi [The Astrophysical Journal Letters, Volume
  848, Issue 2, article id. L27, 9 pp. (2017).] {10.3847/2041-8213/aa90b6}, 848

\bibitem[\protect\citeauthoryear{Tominaga et~al.,}{Tominaga
  et~al.}{2017}]{Tominaga2017}
Tominaga N.,  et~al., 2017, \mn@doi [Publications of the Astronomical Society
  of Japan, Volume 70, Issue 2, id.28] {10.1093/pasj/psy007}, 70

\bibitem[\protect\citeauthoryear{Totani \& Panaitescu}{Totani \&
  Panaitescu}{2002}]{Totani2002}
Totani T.,  Panaitescu A.,  2002, \mn@doi [The Astrophysical Journal, Volume
  576, Issue 1, pp. 120-134.] {10.1086/341738}, 576, 120

\bibitem[\protect\citeauthoryear{Troja et~al.,}{Troja et~al.}{2017}]{Troja2017}
Troja E.,  et~al., 2017, \mn@doi [Nature, Volume 551, Issue 7678, pp. 71-74
  (2017).] {10.1038/nature24290}, 551, 71

\bibitem[\protect\citeauthoryear{Troja et~al.,}{Troja et~al.}{2018}]{Troja2018}
Troja E.,  et~al., 2018, \mn@doi [Monthly Notices of the Royal Astronomical
  Society: Letters] {10.1093/mnrasl/sly061}, 478, L18

\bibitem[\protect\citeauthoryear{Utsumi et~al.,}{Utsumi
  et~al.}{2017}]{Utsumi2017}
Utsumi Y.,  et~al., 2017, \mn@doi [Publications of the Astronomical Society of
  Japan, Volume 69, Issue 6, id.101] {10.1093/pasj/psx118}, 69

\bibitem[\protect\citeauthoryear{Valenti et~al.,}{Valenti
  et~al.}{2017}]{Valenti2017}
Valenti S.,  et~al., 2017, \mn@doi [The Astrophysical Journal Letters, Volume
  848, Issue 2, article id. L24, 6 pp. (2017).] {10.3847/2041-8213/aa8edf}, 848

\bibitem[\protect\citeauthoryear{Villar et~al.,}{Villar
  et~al.}{2017}]{Villar2017}
Villar V.~A.,  et~al., 2017, \mn@doi [The Astrophysical Journal Letters, Volume
  851, Issue 1, article id. L21, 12 pp. (2017).] {10.3847/2041-8213/aa9c84},
  851

\bibitem[\protect\citeauthoryear{Villar et~al.,}{Villar
  et~al.}{2018}]{Villar2018}
Villar V.~A.,  et~al., 2018, \mn@doi [The Astrophysical Journal Letters, Volume
  862, Issue 1, article id. L11, 5 pp. (2018).] {10.3847/2041-8213/aad281}, 862

\bibitem[\protect\citeauthoryear{Wijers \& Galama}{Wijers \&
  Galama}{1999}]{Wijers1999}
Wijers R. A. M.~J.,  Galama T.~J.,  1999, \mn@doi [The Astrophysical Journal,
  Volume 523, Issue 1, pp. 177-186.] {10.1086/307705}, 523, 177

\bibitem[\protect\citeauthoryear{{van Eerten}, {Ryan}, {Ricci}, {Burgess},
  {Wieringa}, {Piro}, {Cenko}  \& {Sakamoto}}{{van Eerten}
  et~al.}{2018}]{vanEerten2018}
{van Eerten} E.~T.~H.,  {Ryan} G.,  {Ricci} R.,  {Burgess} J.~M.,  {Wieringa}
  M.,  {Piro} L.,  {Cenko} S.~B.,   {Sakamoto} T.,  2018, arXiv e-prints, \href
  {http://adsabs.harvard.edu/abs/2018arXiv180806617V} {}

\makeatother
\end{thebibliography}

\appendix
\section{Simulation and post-process calculation}
\label{sec:post_process}
In $ \S $ \ref{sec:analytic_fit} we performed several relativistic hydrodynamic simulations in order to calibrate the analytic expressions in \citet{Nakar2002}.
The full description (physical and grid setups, as well as the post-process calculation) of simulations $ \A, \B $ and $ \C $ is given in \citet{Mooley2018b}.
Note that since the simulations are scale-free, we can determine the absolute values of all parameters by scaling them according to the analytic relations (see \citealt{Granot2012}). In order to compare simulations $ \A, \B $ and $ C $ from \citet{Mooley2018b} we rescaled them to share the same parameters in $ \S $ \ref{sec:analytic_fit}.
For the top-hat simulation we use the RHD module in PLUTO v4.0 \citep{Mignone2007}. We inject a jet with the following parameters: $ \theta_j = 5^\circ, E = 10^{50}\erg $ and initial Lorentz factor $ \Gamma_0 = 200 $, into a uniform medium of a number density $ n = 10^{-3} \cm^{-3} $. Then for the calculation of the synchrotron radiation we assume the standard afterglow model \citep{Sari1997}, and use the numerical post-process calculation method of \citet{Nakar2018}. The grid setup of the top-hat simulation is identical to that of the last phase in simulations $ \A, \B $ and $ C $ in \citet{Mooley2018b}.
 
\label{lastpage}
\end{document}